# Artificial Neural Networks for Detection of Malaria in RBCs

## Purnima Pandit[1], A. Anand[2]

Department of Applied Mathematics[1]
Department of Applied Physics[2]
The Maharaja Sayajirao University of Baroda

E-mail: purnima.pandit-appmath@msubaroda.ac.in

**ABSTRACT:** Malaria is one of the most common diseases caused by mosquitoes and is a great public health problem worldwide. Currently, for malaria diagnosis the standard technique is microscopic examination of a stained blood film. We propose use of Artificial Neural Networks (ANN) for the diagnosis of the disease in the red blood cell. For this purpose features / parameters are computed from the data obtained by the digital holographic images of the blood cells and is given as input to ANN which classifies the cell as the infected one or otherwise.

## Introduction

Malaria is an ancient disease present majorly in the tropical countries having a huge social, economic, and health burden. In recent times, as a result of climatic changes due to global warming, it is predicted to have unexpected effects on Malaria. Both increase and fluctuation in temperature affects the vector and parasite life cycle. An efficient diagnostics is essential for the proper medication and cure. Major clinical diagnostics to identify RBCs affected by malaria is based on microscopic inspection of blood smears, treated with reagents, which stains the malarial parasite. A technician visually inspects these smears to identify malaria-infected red blood cells (RBCs). In developing countries, visual identification of malarial RBCs may become inefficient due to lack of sufficiently trained technicians and poor-quality microscopes and regents.

During such hematologic disorder the changes occur in the structure and viscoelastic properties of individual RBCs. So studying their physical properties can thus contribute greatly to the understanding and possible discovery of new treatments for such diseases. Thus, new advanced mathematical computing techniques are valuable for determining and distinguishing between the healthy and diseased RBCs.

In [1], Anand and at el. have used correlation technique on the images obtained from DHIM to discriminate the malarial cells from the healthy ones. An efficient computational techniques, based on the 3-D images produced by digital holographic interferometric microscopy (DHIM) [6], to automatically discriminate between malarial and healthy RBCs can be very beneficial. It will be advantageous if such DHIM capturing instruments are portable and easy to use. We propose the technique to distinguish the malaria infected RBCs from the healthy ones using Artificial Neural Network (ANN), based on its physical/statistical features that are extracted from the images obtained from Digital Holographic Interferometric Microscope (DHM). Artificial Neural Network (ANN) for malaria diagnosis is one time trained intelligent program which is easy to use, portable, low cost and make malaria diagnosis more rapid and accurate.

# DIGITAL HOLOGRAPHIC INTERFEROMETRIC MICROSCOPY

A hologram is the photographically or otherwise recorded interference pattern between a wave field scattered from the object and a coherent background, called the reference wave. The digital holographic interferometric microscopy, comparisons of phases at the image plane, obtained from the recorded holograms with the object present and with just the background, provides object phase information. This can be converted into thickness information of cells. Then efficient computational techniques will be applied on the 3-D images produced by DHM for micro objects, to automatically discriminate between malarial and healthy RBCs.

Dennis Gabor invented holography in 1948 as a method for recording and reconstructing the amplitude and phase of a wavefield. The word *holography* is derived from the Greek words 'holos' meaning 'whole' or 'entire' and 'graphein' meaning 'to write'. A hologram is the photographically or otherwise recorded interference pattern between a wave field scattered from the object and a coherent background, called the reference wave.

## 3.1. Digital Hologram

The general set-up for recording off-axis holograms is shown in figure 1. Light with sufficient coherence length is split into two partial waves by a beam splitter (BS). One wave illuminates the object, is scattered and reflected to the recording medium, e.g. a photographic plate. The second wave, called the reference wave, illuminates the plate directly. Both waves are interfering. The interference pattern is recorded, e.g. by chemical development of the photographic plate. The recorded interference pattern is called a hologram.

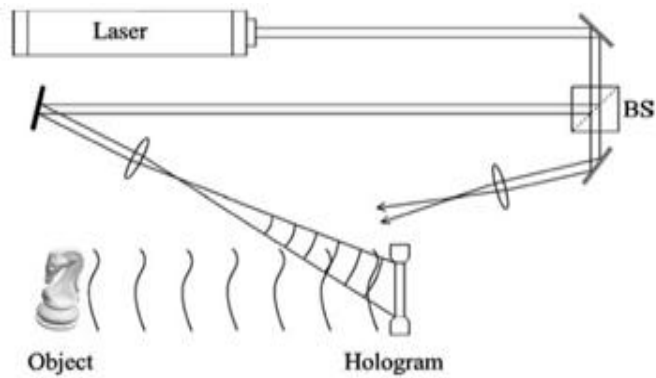

**Fig. 1: Hologram Recording**

The original object wave is reconstructed by illuminating the hologram with the reference wave, figure 2. An observer sees a virtual image, which is indistinguishable from the image of the original object. The reconstructed image exhibits all the effects of perspective and depth of focus.

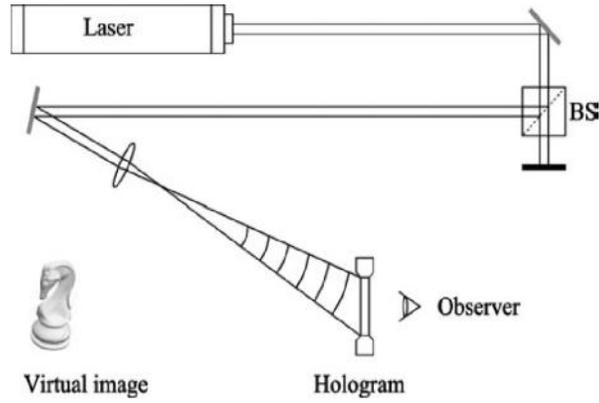

Fig. 2: Hologram reconstruction

The holographic process is described mathematically as follows:

$$O(x,y) = o(x,y) \exp(i\varphi_o(x,y)) \qquad (1)$$

Equation (1) the complex amplitude of the object wave with real amplitude $o$ and phase $\varphi_o$ and

$$R(x,y) = r(x,y)\exp(i\varphi_R(x,y)) \qquad (2)$$

Equation (2) gives the complex amplitude of the reference wave with real amplitude $r$ and phase $\varphi_R$. Both waves interfere at the surface of the recording medium. The intensity is calculated by

$$I(x,y) = |O(x,y) + R(x,y)|^2$$

$$= (O(x,y) + R(x,y))(O(x,y) + R(x,y))^*$$

$$= R(x,y)R^*(x,y) + O(x,y)O^*(x,y) + O(x,y)R^*(x,y) + R(x,y)O^*(x,y)$$

where, * denotes the conjugate complex.

The amplitude transmission $h(x, y)$ of the developed photographic plate (or any other recording media) is proportional to $I(x, y)$, that is,

$$h(x, y) = h_0 + \tau I(x, y)$$

where is a constant, $\tau$ is the exposure time and $h_0$ is the amplitude transmission of the unexposed plate. $h(x, y)$ is also called the hologram function. In digital holography using charge-coupled device detector CCDs as recording medium $h_0$ can be neglected.

For hologram reconstruction the amplitude transmission has to be multiplied with the complex amplitude of the reconstruction (reference) wave:

$$R(x,y)h(x,y) = [h_0 + \tau(r^2 + o^2)]R(x,y) + \tau r^2 O(x,y) + \tau R^2(x,y)O^*(x,y)$$

The first term on the right side of this equation is the reference wave, multiplied by a factor. It represents the undiffracted wave passing through the hologram (zero diffraction order).

The second term is the reconstructed object wave, forming the virtual image. The factor $\tau r^2$ only influences the brightness of the image. The third term produces a distorted real image of the object. For off-axis holography the virtual image, the real image and the undiffracted wave are spatially separated.

### 3.1.1 Recording in Digital holography

In a set-up for digital recording of off-axis holograms is shown in figure 3. A plane reference wave and the wave reflected from the object are interfering at the surface of a CCD. The resulting hologram is electronically recorded and stored. The object is, in general, a three-dimensional body with diffusely reflecting surface, located at a distance $d$ from the CCD.

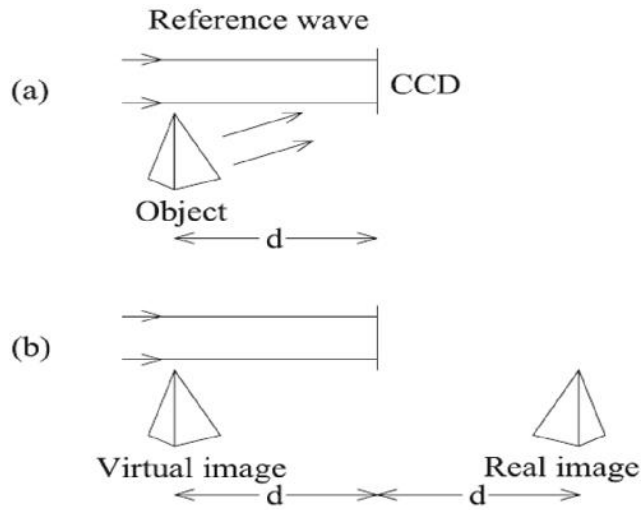

**Fig.3: Digital holography: (a) recording, (b) reconstruction.**

In optical reconstruction the virtual image appears at the position of the original object and the real image is formed also at a distance $d$, but in the opposite direction from the CCD, see figure 3(b).

The diffraction of a light wave at an aperture (in this case a hologram) which is fastened perpendicular to the incoming beam is described by the Fresnel–Kirchhoff integral

$$\Gamma(\xi,\eta) = \frac{i}{\lambda}\int_{-\infty}^{\infty}\int_{-\infty}^{\infty} h(x,y)R(x,y)\frac{\exp(-i\frac{2\pi}{\lambda}\rho)}{\rho} \times \left(\frac{1}{2}+\frac{1}{2}\cos\theta\right) dx\, dy \qquad (3)$$

with

$$\rho = \sqrt{(x-\xi)^2+(y-\eta)^2 + d^2} \qquad (4)$$

where, is the distance between a point in the hologram plane and a point in the reconstruction plane, see figure 4. The angle is also defined in figure 4. The diffraction pattern is calculated at a distance $d$ behind the CCD plane, which means it reconstructs the complex amplitude in the plane of the real image.

Equation (3) is the basis for numerical hologram reconstruction. Because the reconstructed wave field ( , ) is a complex function, both the intensity as well as the phase can be calculated. This is in contrast to the case of optical hologram reconstruction, in which only the intensity is made visible.

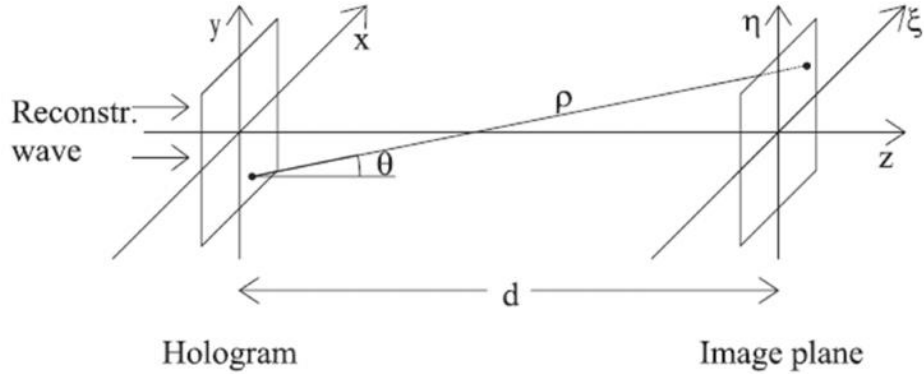

**Fig. 4: Coordinate system**

The set-up of figure 4 is often used in digital holography, because a plane wave propagating perpendicularly to the surface of the CCD can be easily arranged in the laboratory. Another advantage is that for this geometry the reconstructed real image has no geometrical distortions.

**3.1.2 Reconstruction by the Fresnel transformation**

For $x$ and $y$ values as well as for   and   values which are small compared to the distance $d$ between the reconstruction plane and the CCD expression (4) can be replaced by the first terms of the Taylor series:

$$\rho = d + \frac{(\xi - x)^2}{2d} + \frac{(\eta - x)^2}{2d} - \frac{1}{8}\frac{[(\xi - x)^2(\eta - x)^2]}{d^3} + \ldots$$

$$\approx d + \frac{(\xi - x)^2}{2d} + \frac{(\eta - x)^2}{2d}$$

with the additional approximation $\cos\theta \approx 1$.

By replacing the dominator in (3) by $d$ the following expression results:

$$\Gamma(\xi, \eta) = \frac{i}{\lambda d} \exp\left(-i\frac{2\pi}{\lambda}d\right) \int_{-\infty}^{\infty}\int_{-\infty}^{\infty} R(x, y) h(x, y)$$
$$\times \exp\left[(-i\frac{\pi}{\lambda d})((\xi - x)^2 + (\eta - y)^2)\right] dx\, dy$$

If we carry out the multiplications in the argument of the exponential under the integral we get

$$\Gamma(\xi,\eta) = \frac{i}{\lambda d}\exp\left(-i\frac{2\pi}{\lambda}d\right)\exp\left[\left(-i\frac{\pi}{\lambda d}\right)(\xi^2+\eta^2)\right]$$
$$\times \int_{-\infty}^{\infty}\int_{-\infty}^{\infty} R(x,y)h(x,y)\exp\left[\left(-i\frac{\pi}{\lambda d}\right)(x^2+y^2)\right]$$
$$\times \exp\left[i\frac{2\pi}{\lambda d}(x\xi+y\eta)\right] dx\, dy$$

This equation is called the *Fresnel approximation* or *Fresnel transformation*. It enables reconstruction of the wave field in a plane behind the hologram, in this case in the plane of the real image.

The intensity is calculated by squaring

$$I(\xi,\eta) = |\Gamma(\xi,\eta)|^2$$

The phase is calculated by

$$\varphi(\xi,\eta) = arctan\frac{Im[\Gamma(\xi,\eta)]}{Re[\Gamma(\xi,\eta)]}$$

where, *Re* denotes the real part and *Im* the imaginary part. These phase values are used for further processing of the DHMI of the RBCs.

## Artificial Neural Networks

Artificial Neural Networks (ANN) are massively parallel, distributed processing systems representing a computational technology built on the analogy to the human information processing system. They are massively connected networks of simple processing elements called neurons. They have a natural propensity to learn and save the knowledge to make it available for use. ANNs can be used for classification, pattern recognition and function approximation.

It is an information processing system consisting of large number of interconnected processing units. It works in a way our nervous system processes the information. It is basically a dense interconnection of simple non-linear computational elements. It has been successfully applied to problems involving pattern classification, function approximation, regression, prediction and others.

The biological aspects of neural networks were modeled by McCullough and Pitts in 1943. This model exhibits all the properties of neural elements like excitation potential thresholds for neuron firing and non-linear amplification which compresses the strong input signal. Various kinds of Neural Networks are used for different purpose. Single-layer Perceptron, Multi-layer Perceptron, Hopfield recurrent networks, Kohonnen self-organizing networks, Radial Basis Function Networks, Support Vector Machines and Extreme Learning Machines are most common amongst them.

The fundamental element of neural network is neuron. It is an operator, which maps $\mathbb{R}^n \to \mathbb{R}$ and it is explicitly described by the equation.

$$y_j = f\left(\sum_{i=1}^{n} w_{ji} x_i + w_0\right) \qquad (5)$$

Here, $X^T = [x_1, x_2, \ldots, x_n]$ is an input vector, $W_j^T = [w_{j1}, w_{j2}, \ldots, w_{jn}]$ is the weight vector to the $j^{th}$ neuron and $w_0$ is the bias. $f: \mathbb{R} \to (-1,1)$ is a monotonic, continuous function such as $tanh(x)$ or $signum(x)$. Such neurons are interconnected to form a network. Amongst various activation functions, the sigmoid activation function is more useful since it is continuous and differentiable. The simple neural network is as shown in Fig. 5.

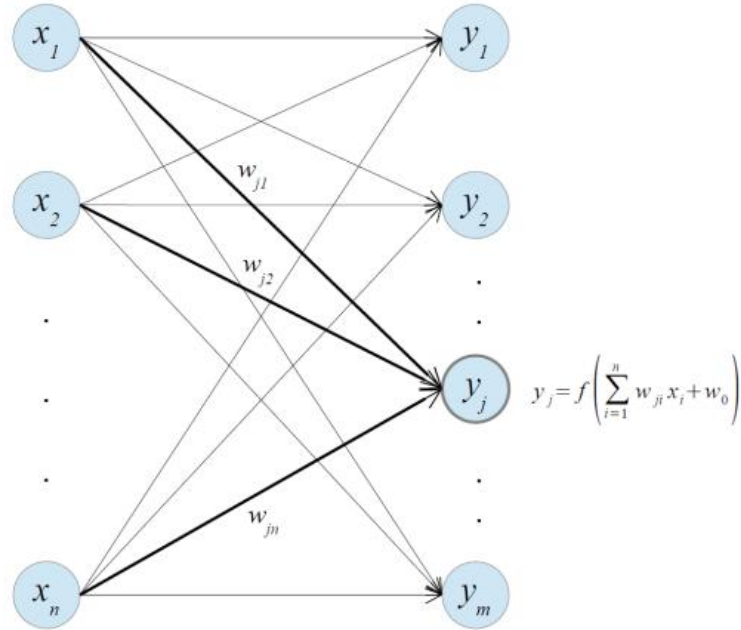

**Fig. 5: Feed forward Neural Network**

The Networks are categorized into mainly two categories: Feed Forward and Recurrent. The learning algorithms are categorized in three types: supervised learning, unsupervised learning and reinforced learning. We have used multi-layer network Radial Basis Function Network, which is a supervised learning algorithm.

Learning (or Training) is a process by which a network adapts changes by adjusting its changeable parameters i.e. weights. They change in such a way that there is an association between set of inputs and desired outputs. In Artificial Neural Networks, The general rule for weight updating is $W_{k+1} = W_k + \Delta W_k$. Here, $W_{k+1}$ is weight in $(k+1)^{st}$ iteration, $W_k$ is the weight in $k^{th}$ iteration and $\Delta W_k$ is change in weight in $k^{th}$ step, which can be determined using different algorithms.

ANN plays a primary role in contemporary artificial intelligence and machine learning. The use of ANN in function approximation resulted from the following facts:

(i) Cybenko and Hornik proved that the multi-layered Neural Network is universal approximator. It can approximate any continuous function defined on compact set.

(ii) The Back-propagation algorithm used for the training of the feed-forward Neural Networks with the hidden layers.

The Multi-layer feed forward network (MFFN) trained with Back Propagation Rule is one of the most important neural network model. In the real practical situations when there is the lack of the mathematical model for the system, the ANN can be trained with the help of Input-Output pairs without fitting any kind of model to the system.

The fundamental element of neural network is neuron. We can refer to neuron as an operator, which maps $\mathbb{R}^n \to \mathbb{R}$ and is explicitly, described by the equation (6).

$$x_j = \Gamma\left(\sum_{i=1}^{n} w_{ji} u_i + w_0\right) \tag{6}$$

Here, $U^T = [u_1, u_2, \ldots, u_n]$ is input vector, $W_j^T = [w_{j1}, w_{j2}, \ldots, w_{jn}]$ is weight vector to the $j^{th}$ neuron and $w_0$ is the bias. $\Gamma(\cdot)$ is a monotone continuous function, $\Gamma: \mathbb{R} \to (-1,1)$ (e.g. $tanh(\cdot)$ or $signum(\cdot)$). Such neurons are interconnected to form a network.

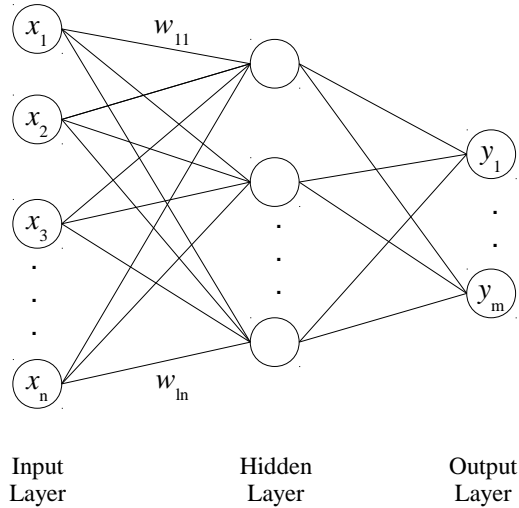

Input Layer     Hidden Layer     Output Layer

**Fig. 6: A Multi Layered Feed-Forward Network**

In the Feed-forward network, neurons are organized in layers $h = 0, 1, \ldots, L$. It is common practice to refer to the layer $h = 0$ as the input layer and $h = L$ as the output layer and to all other as hidden layers. A neuron at layer $h$ receives its inputs only from neurons in the $h - 1$ layer. The output of the $i^{th}$ element in the layer $h$ is given by

$$y_i^h = \Gamma\left(\sum_j w_{i,j}^h y_j^{h-1} + w_{i,0}^h\right) \tag{7}$$

Here, $\left[w_i^h\right]^T = [w_{i,0}^h, w_{i,1}^h, \ldots, w_{i,n_{L-1}}^h]$ is the weight vector associated with $i^{th}$ neuron at the $h^{th}$ layer. Such a family of networks with $n_i$ neurons at the $i^{th}$ layer will be denoted by $N_{n_0,n_1,\ldots,n_L}^L$.

The general learning rule for weight updating is $W(k+1) = W(k) + \Delta W(k)$, where, $W(k+1)$ is weight in $(k+1)^{th}$ step, $W(k)$ is weight in $(k)^{th}$ step and $\Delta W(k)$ is change in weight in $(k)^{th}$ step. The change in weight for supervised delta rule is given by

$$\Delta w_{lj} = \eta f'\left(\sum_{i=0}^{n} w_{ji} x_i\right) f\left(\sum_{i=0}^{n} w_{ji} x_i\right) \tag{8}$$

The training in the Feed-forward type Multilayered neural network is done using back-propagation algorithm. Back-propagation algorithm is an extension of delta rule for training for multilayer feed-forward networks.

## Experimental Work

We have the holographic image data for 48 RBC, which are processed and then various features are obtained in following manner. The phase data for the cells are obtained in gray scale to which thresholding or masking is done for the selection of portion of the image to work with.

The images obtained after converting it into grey scale for image of edge for different mask values 0.28 and 0.3 is shown in following figures:

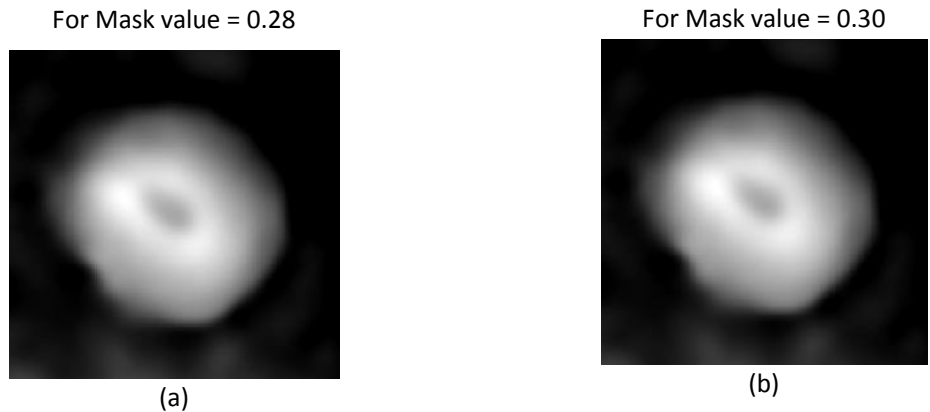

For Mask value = 0.28     For Mask value = 0.30

(a)     (b)

**Fig. 7: Image obtained after converting to black and white**

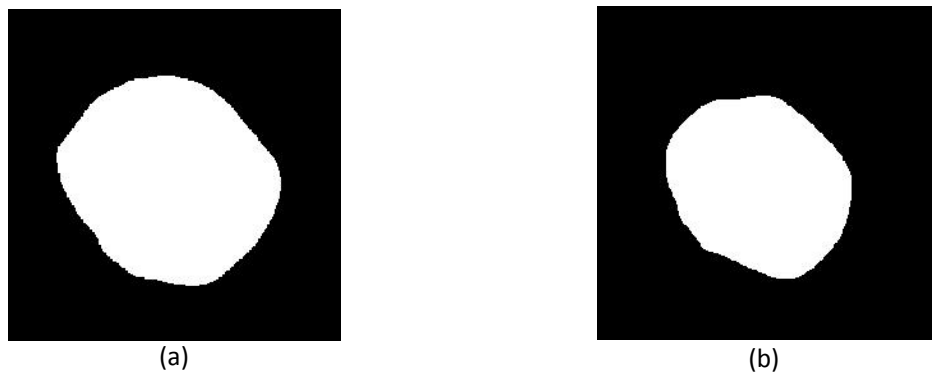

(a)     (b)

**Fig. 8: Image obtained after converting into binary, to find area of RBC**

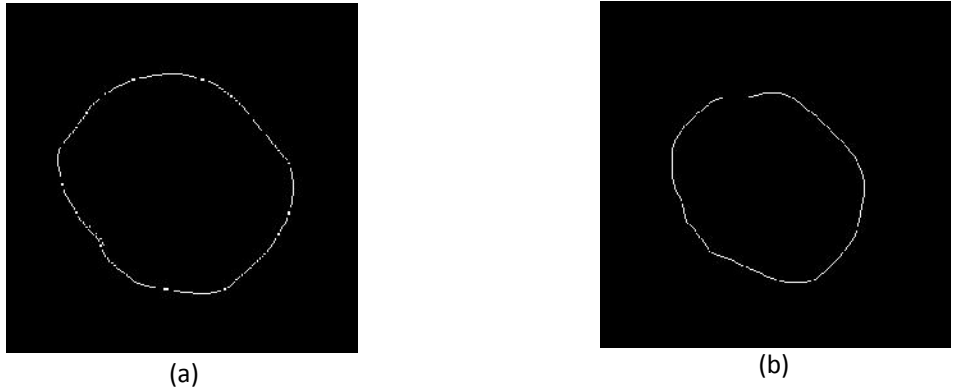

(a)            (b)

**Fig. 9: Image obtained after detecting edges, to find perimeter of RBC**

Once the portion of RBC image was selected the parameters were calculated for it like taking average of the phase of the image, standard deviation and coefficient of variation. To find the area we count the number of pixels selected for processing and the border pixels to obtain the perimeter. Circularity is calculated using the formula

$$c = \frac{4f\,A}{p^2}$$

Ideally, healthy RBC is doughnut shaped and infected RBC becomes spherical due to growth of parasites within the RBC. In our work the detection of healthy and infected RBC is done using six parameters. Table 1. and Table 2., shows such calculated values of six parameters: area, perimeter, circularity, mean, standard deviation, and covariance. These parameters were obtained using holographic images of 24 healthy RBC and 24 infected RBC.

| | Parameters of 24 Healthy cells | | | | | |
|---|---|---|---|---|---|---|
| | Area | Perimeter | Circularity | Mean | Std | Cov |
| 1 | 32983 | 589 | 1.0930 | 0.6479 | 0.8203 | 0.6729 |
| 2 | 29156 | 560 | 1.0809 | 0.5883 | 0.8281 | 0.6858 |
| 3 | 22000 | 515 | 1.0210 | 0.5446 | 0.8771 | 0.7693 |
| ⋮ | | | | | | |
| 24 | 19314 | 475 | 1.0372 | 0.4604 | 0.7787 | 0.6064 |

**Table 1: Parameter values for Healthy cells**

| | Parameters of 24 Malarial cells | | | | | |
|---|---|---|---|---|---|---|
| | Area | Perimeter | Circularity | Mean | Std | Cov |
| 1 | 23987 | 558 | 0.9839 | 0.5146 | 0.8028 | 0.6445 |
| 2 | 26365 | 534 | 1.0779 | 0.5585 | 0.7518 | 0.5651 |
| 3 | 22401 | 480 | 1.1053 | 0.4538 | 0.7012 | 0.4916 |
| ⋮ | | | | | | |
| 24 | 22640 | 498 | 1.0711 | 0.3867 | 0.6574 | 0.4322 |

**Table 2: Parameter values for Malaria infected cells**

The six parameters obtained after preprocessing on holographic images are used as inputs for the detection of malaria infected and healthy cells to the Artificial Neural Network (ANN). The outputs of ANN are binary bipolar, -1 stands for healthy and 1 stands for infected RBC.

MATLAB is used to create Feed-forward neural network. The training was done using Back-propagation. We had 48 patter10s consisting of 24 healthy and 24 malarial cells, out of which 5 healthy and 5 malarial random patterns are kept for testing and remaining 38 patterns are used for training purpose. We considered an ANN with architecture, $N_{6,20,50,30,1}^5$, this network got trained in 120 iterations as shown in Figure 10., the error convergence and the state performance as shown in Figure 11.

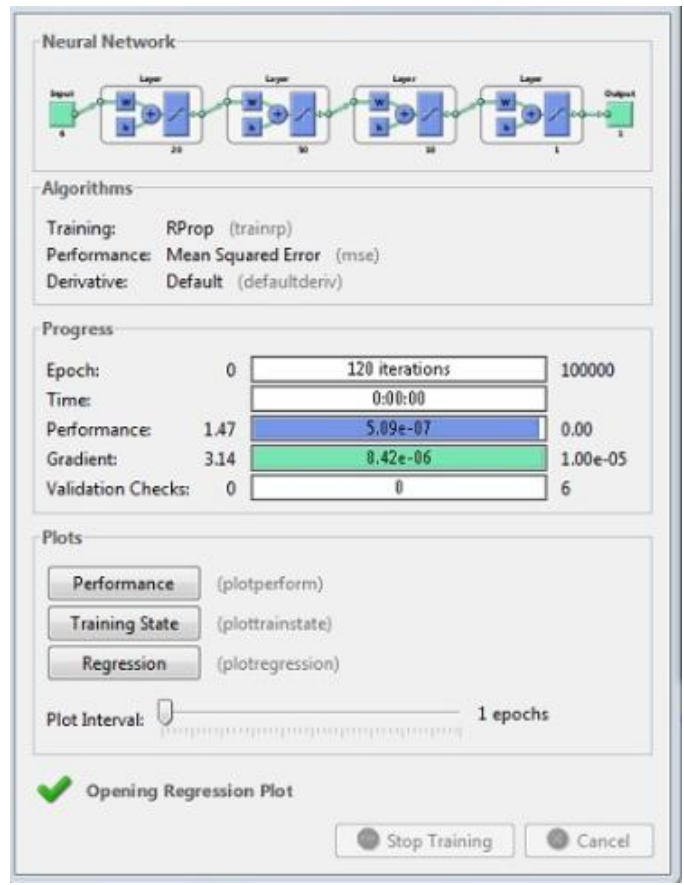

**Fig. 10: Trained ANN Window**

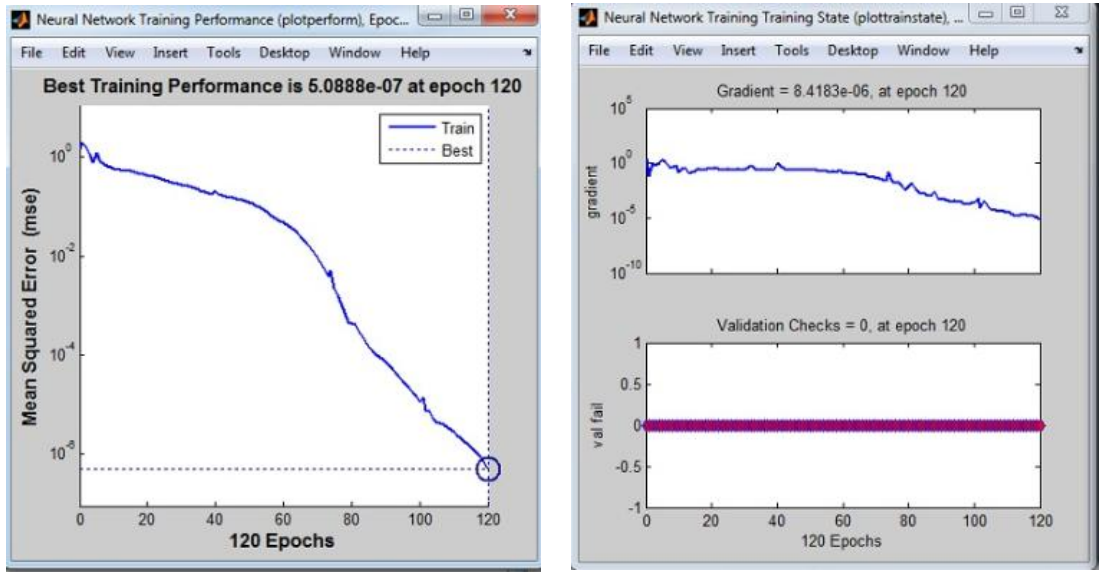

**Fig. 11: Training Performance and Training state**

This trained network is the tested for the remaining testing patterns. The simulation results for the trained as well as testing inputs is as shown in the Figures 12 (a) and (b). The 90% results for the testing inputs are encouraging and justify the use of ANN for detection of the disease.

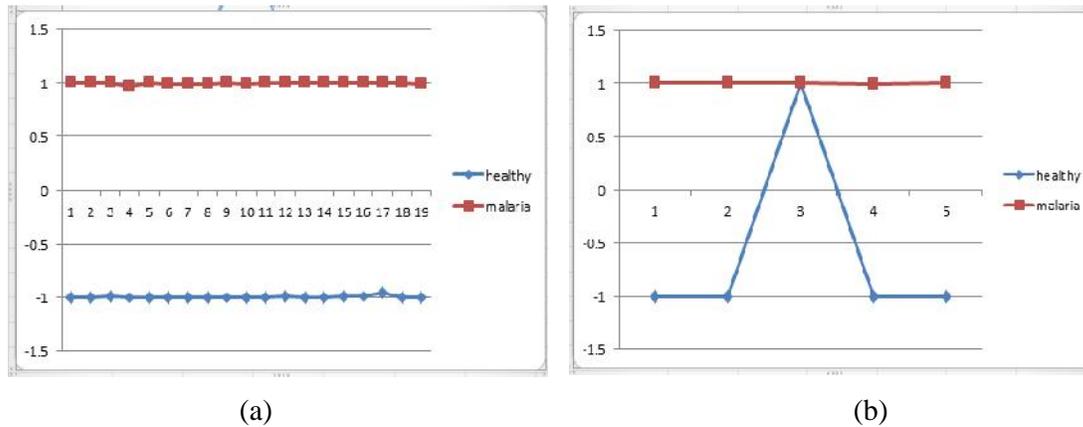

(a)                                                                 (b)

**Fig. 12: The classification by ANN for the: (a) training and (b) testing patterns**

## Conclusion

We have determined various parameters by processing the holographic images of healthy as well as infected RBCs. We have used Artificial Neural Network (ANN) for the detection of malaria infected RBCs. In the proposed techniques for the detection of malaria infected RBCs done using ANN gives results almost from 90-100%. These results can be refined more by increasing the number of parameters and also the sample size of the data provided.

*Acknowledgement:* The work is supported by RCC-MSU (RCC/Dir./2015/177) Research fund.